\documentclass{article}

\usepackage[utf8x]{inputenc}

\usepackage{textcomp,marvosym}

\usepackage{fixltx2e}

\usepackage{amsmath,amssymb}

\usepackage{cite}

\usepackage{nameref}

\usepackage{hyperref}

\usepackage{breakurl}

\usepackage{caption}

\makeatletter
\renewcommand{\@biblabel}[1]{\quad#1.}
\makeatother

\date{}

\usepackage{lastpage,fancyhdr,graphicx}
\usepackage{epstopdf}

\begin{document}
\vspace*{0.2in}

\begin{flushleft}
{\Large
\textbf\newline{SensibleSleep: A Bayesian Model for Learning Sleep Patterns from Smartphone Events} 
}
\newline
\\
Andrea Cuttone\textsuperscript{1\Yinyang*},
Per B{\ae}kgaard\textsuperscript{1\Yinyang},
Vedran Sekara\textsuperscript{1,3},
H{\aa}kan Jonsson\textsuperscript{3},
Jakob Eg Larsen\textsuperscript{1},
Sune Lehmann\textsuperscript{1,2}
\\
\bigskip
\textbf{1} DTU Compute, Technical University of Denmark, Kgs. Lyngby, Denmark
\\
\textbf{2} The Niels Bohr Institute, University of Copenhagen, Copenhagen, Denmark
\\
\textbf{3} Sony Mobile, Nya Vattentornet, Mobilv{\"a}gen, 221 88 Lund, Sweden.
\\
\bigskip

\Yinyang These authors contributed equally to this work.

* ancu@dtu.dk

\end{flushleft}

\section*{Abstract}
We propose a Bayesian model for extracting sleep patterns from smartphone events.
Our method is able to identify individuals’ daily sleep periods and their evolution over time, and provides an estimation of the probability of sleep and wake transitions.
The model is fitted to more than $400$ participants from two different datasets, and we verify the results against ground truth from dedicated armband sleep trackers. 
We show that the model is able to produce reliable sleep estimates with an accuracy of 0.89, both at the individual and at the collective level.
Moreover the Bayesian model is able to quantify uncertainty and encode prior knowledge about sleep patterns.
Compared with existing smartphone-based systems, our method requires only screen on/off events, and is therefore much less intrusive in terms of privacy and more battery-efficient.

\section*{Introduction}

Sleep is an important part of life, and quality of sleep has a significant impact on individual well-being and performance.
This calls for methods to analyze sleep patterns in large populations, preferably without laborious or invasive consequences, as people typically disapprove of the use of intrusive technologies~\cite{Choe:2011:OCT:1978942.1979395}.

Large scale studies of human sleep patterns are typically carried out using questionnaires, a method that is known to be unreliable.
It is possible to perform more accurate studies, but these are currently carried out within small controlled environments, such as sleep labs.
In order to perform accurate measurements of sleep in large populations---consisting of thousands of individuals---without dramatically increasing costs, alternative methods are needed.

Smartphones have become excellent proxies for studies of human behavior \cite{candia2008uncovering, gonzalez2008understanding}, as they are able to automatically log data from built-in sensors (GPS, Bluetooth, WiFi) and on usage patterns (phone calls, SMS and screen interaction), from which underlying user behavioral patterns can be derived.

Smartphone data has been used to infer facets of human behavior such as social interactions \cite{sekara2015fundamental}, communication \cite{onnela2007structure}, mobility \cite{zheng2009mining}, depression \cite{saeb2015mobile} and also sleep patterns \cite{abdullah2014towards}.
Either paired with additional sensors or on their own, mobile app solutions are able -- sometimes very ingeniously -- to track individual sleep patterns and visualize them.
We cite as examples \emph{Smart Alarm Clock}~\cite{smart-alarm-clock}, \emph{Sleep Cycle}~\cite{sleep-cycle}, \emph{SleepBot}~\cite{sleepbot}, and \emph{Sleep as Android}~\cite{sleep-as-android}.

Using mobile phone data to derive sleep patterns has thus already been demonstrated and verified, and offers advantages (i.e. reduced cost) as an alternative to dedicated sleep monitoring devices.
In this paper we suggest extending previous approaches, using a Bayesian model to infer rest and wake periods based on smartphone screen activity information.
The advantages of our proposed Bayesian approach \textit{SensibleSleep}, as compared to previous work, are that it:
\begin {itemize}
\item is less sensitive to ``noisy'' data, for instance infrequent phone usage during sleep interruptions (such as checking the phone at night)
\item is able to quantify not only specific rest and wake times but also characterize their distributions and thus uncertainty
\item can encode specific prior beliefs, for instance on expected rest periods (when desirable)
\item can capture complex dependencies between model variables, and possibly even detect and relate patterns that are common to a group of people with diverging individual patterns (when using one of the proposed hierarchical models), such as detecting how available daylight may modulate sleep patterns across an otherwise heterogeneous group of users
\end {itemize}
Our method, moreover, only needs screen on/off events and is thus \emph{non-intrusive}, \emph{privacy-preserving}, and has \emph{lower battery cost} than microphone or ac\-celerom\-e\-ter based ones.

We start by providing an overview of the related work.
We then describe the collected data, and introduce the Bayesian model.
We compare the model results with ground truth obtained by sleep trackers, and show how the model is able to infer the sleep patterns with high accuracy.
Finally we describe the individual and collective sleep patterns inferred from the data.

\section*{Related Work}
A key finding by Zhang et al.~\cite{zhang2015global} shows a global prevalence of sleep deprivation in a group of students, partly linked to heavy media usage.
In this study sleep patterns are largely deduced from the teachers' perception or based on individual self-reports, lacking more direct measurements.

Corroborating this finding, Orzech et al.~\cite{orzech2016digital} report that digital media usage before bedtime is common among university students, and negatively impacts sleep. 
The findings are based on studies involving self-reports through (online) sleep diaries and digital media surveys, and also lacks more direct measurements of sleep patterns. 
Additionally, this would make it possible to increase the scale of the experiment and enable the study of larger populations.

Abdullah et al.~\cite{abdullah2014towards} have previously demonstrated using 9 subjects how a simple rule-based algorithm is able to infer sleep onset, duration and midpoint based on a (filtered) list of screen on-off patterns with the help of previously learned individual corrective terms, and further analyzed behavioral traits of the inferred \emph{circadian rhythm}~\cite{richter1960biological,aschoff1965circadian}. 
The algorithm uses an initial two weeks of data with journal self-reported sleep for learning key corrective terms in order to improve the accuracy and compensate for differences between \emph{actual} sleep and  \emph{inferred} nightly rest period.
The method has been verified against a daily online sleep journal and results in differences less than 45 minutes of average sleep duration over the entire analysed period.
While our proposed Bayesian model, which has been applied to more than 400 users, may be more complex, it increases the robustness and allows us to better quantify the uncertainties of the inferred resting periods as well as offer the possibility of building more advanced models across heterogeneous groups of users. 
In particular, our model may better be able to handle short midnight interruptions, which appear to be not uncommon, without any additional filtering.

In contrast to Abdullah et al. using (only) screen on-off events, a fine-grained sleep monitoring by ``hearing'' and analyzing breathing through the earphone of a smartphone is suggested by Ren et al.~\cite{ren2015fine}.
Here six users tested the system over a period of 6 months, demonstrating the feasibility of using smartphones for the purpose of analysing breathing patterns, using a Respiration Monitor Logger as ground truth.
Sleep estimates are not directly inferred in this paper, however.
This technology is also non-invasive, although it does requires capturing and analyzing large samples of audio data.

\emph{iSleep}~\cite{hao2013isleep} proposes detecting sleep patterns by means of a decision tree model, also based on audio features.
The system was evaluated with 7 users for a total of 51 days, and shows high accuracy in detecting snoring and coughing as well as sleep periods, but report drops in performance due to ambient noise.

Increasing the number of features, the \emph{Best Effort Sleep model}~\cite{chen2013unobtrusive} is based on a linear combination of phone usage, accelerometer, audio, light, and time features using a self-reporting sleep journal, and subsequently achieved a 42 minutes mean error on 8 subjects in a test period of 7 days.

Other work also tries to estimate sleep quality, for example \emph{Intelligent Sleep Stage Mining Service with Smartphones}~\cite{Gu:2014:ISS:2632048.2632084}, which uses Conditional Random Fields on a similar set of features trained on 45 subjects over 2 nights, and reports over 65\% accuracy of detection of sleep phases, compared to EEG ground truth on 15 test subjects over 2 nights.

\emph{Candy Crushing Your Sleep}~\cite{Jayarajah:2015:CCY:2800835.2804393} uses the long\-est period of phone usage inactivity as heuristic for sleep, with some ad-hoc rules for merging multiple periods, and proceeds to quantify the sleep quality and to identify aspects of daily life that may affect sleep.
The inferred sleep period was however not validated against any ground truth.

The \emph{Sleep Well} framework~\cite{Hossain:2015:SWS:2847603.2848206} deploys a Bayesian probabilistic change-point detection, in parallel with an unsupervised classification, of features extracted from accelerometer data, in order to identify fine-grained sleep state transitions.
It then uses an active learning process to allow users to incrementally label sleep states, improving accuracy over time.
It was evaluated both on existing datasets with clinical ground truth, and on 17 users for 8-10 days with user diary data as ground truth, reaching an average sleep stage classification accuracy approaching 79\%.

In comparison, even though sleep quality is not estimated, our non-intrusive model only needs screen on/off events and has been tested on a large user-base, and can suitable for very large-scale deployment.

\section*{Methods}

\subsection*{Data Collection}
We have analyzed two datasets in this work.

The first dataset (A) was provided by Sony Mobile, and contains smartphone app launches coupled with sleep tracking data from the SWR10 and SWR30 fitness tracking armbands~\cite{sony-smartwear}.
For each user we have a set of records containing an anonymized unique user identifier, a timestamp and the unique app package name.
Note that the model only uses the app launch timestamp and completely ignores the app identifier, therefore no privacy risks related to app names are present.
The sleep tracking data indicates when each user is detected asleep or awake with a granularity of one minute, serving as ground truth that we will compare our results against.
From this dataset we select 126 users that have at least 3 hours of tracked sleep per day, and have between 2 and 4 weeks of contiguously tracked sleep.

The second dataset (B) originates from the \emph{SensibleDTU} project \cite{stopczynski2014measuring}, which collected smartphone sensor data for more than 800 students at the Technical University of Denmark.
In this dataset we focus on the screen interaction sensor that records whenever the smartphone screen is turned on or off, either by user interaction or by notifications.
Each record contains a unique user identifier, a timestamp, and the event type (on or off).
From this dataset we select \(324\) users in November 2013 that have at least 10 events per day, thus filtering out users with gaps in the collected data or with very sparse data.
There is on average $\approx 76$ screen-on activations pr. day pr. user in this period.

Data collection for the SensibleDTU dataset was approved by the Danish Data Protection Agency, and informed consent has been obtained for all study all participants.
Data collection for the Sony dataset has been approved by the Sony Mobile Logging Board and informed consent has been obtained for all study participants according to the Sony Mobile Application Terms of Service and the Sony Mobile Privacy Policy.

\subsection*{Model Assumptions}

The underlying assumptions of the model are (1) that the user is in one of two modes: being \emph{awake} or \emph{sleeping}, and (2) that mobile phone usage differs between the two modes.
In particular a user will have many screen interactions when awake, and very few or even no interactions when sleeping.

Sleeping is here considered as an extended \emph{resting} period that typically takes place once every 24 hours at roughly similar times, as governed by the users circadian rhythm and influenced by socio-dynamic structures, during which the owner physically rests and/or sleeps. 
Resting periods, however, might be interrupted by short periods of activity, such as checking the time on the phone or responding to urgent messages. 
This behavior leads to two different activity levels, which we label $\lambda_{awake}$ and  $\lambda_{sleep}$, one for each mode. 

If we can deduce when the switchpoint between the two distributions occur during each \(24\) hour period, we can also infer the time during which the owner is \emph{resting} for the night, and thereby also the period within which sleeping takes place.

Short of using the more invasive EEG or polysomnographic methods, properly differentiating the resting period and actual sleep is difficult; even sleep diaries may easily contain reporting bias or be somewhat inaccurate. 
To remove self-reporting bias and to study a larger population we have therefore decided on using a motion-based detector (Sony fitness tracking armbands) as ground truth.

If higher accuracy would be required, applying individual corrective terms (i.e. average sleep/rest time differences) learned from an initial period by more accurate means (polysomnography, external observer or possibly a careful user diary) might be possible, similar to what as demonstrated by Abdullah et al.~\cite{abdullah2014towards}.

\subsection*{Model Structure}

Each user is considered independently. 
We divide time into $24-$hour periods starting at 16:00 and ending at 15:59 on the next calendar day, so that the night period and the expected sleep midpoint is in the middle, for convenience.
Each day is divided into $n=24*4=96$ time bins of size $15$ minutes.
We count the number of events that start within each time bin, where an event is an app launch for dataset A and a screen-on for dataset B.
Information about the duration of the events is purposely discarded, as phone usage typically takes place in short bursts.
This is supported by the median duration of screen events in dataset B, which is $\approx 26.5$ seconds.
It is reasonable to assume that the count of events $k$ in each time bin follows a Poisson distribution:

\[P(k) = \operatorname{Poisson}(k,\lambda)= {\frac {\lambda ^k e^{{-\lambda }}}{k!}}\]
\\
with $\lambda = \lambda_{awake}$ or $\lambda = \lambda_{sleep}$, depending on the mode of the user.
It is, furthermore, assumed that the user mode, and consequently the value for $\lambda$, is determined by two switchpoint variables $t_{sleep}$ and $t_{awake}$, both assuming values from $0$ to $n$:

\begin{equation*}
\lambda = 
\begin{cases}
       \lambda_{sleep} & \operatorname{if} t_{sleep} \leq t < t_{awake}\\
       \lambda_{awake} & \operatorname{if} t < t_{sleep} \lor t \geq t_{awake}\\
\end{cases}
\end{equation*}
For simplicity, all models assume that $\lambda_{sleep}$ is identical for all days of a given user.
It can be expected that users have a very low number of screen events during sleep mode, which is encoded in this prior belief:

\begin{equation*}
\lambda_{sleep} \sim \operatorname{Exponential}(10^4)
\end{equation*}
Here $\operatorname{Exponential}$ represents the exponential distribution: 

\begin{equation*}
f(x;\lambda) = \begin{cases}
\lambda e^{-\lambda x} & x \ge 0\\
0 & x < 0
\end{cases}
\end{equation*}
The rate parameter is set to a very large value to encode our prior belief that almost no events should happen during the sleep time.

\begin{figure}[!h]
\centering
\includegraphics[width=1\columnwidth]{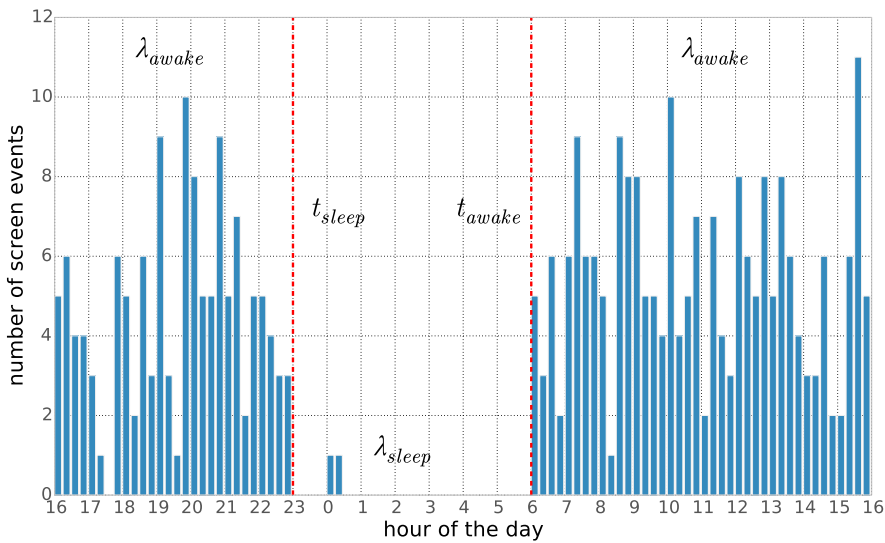}
\caption{Conceptual illustration of the model. 
We assume that for each day the event counts follow two different Poisson distributions: one for sleep periods (rate $\lambda_{sleep}$) and one for awake periods (rate $\lambda_{awake}$).
Furthermore we assume that two switchpoints $t_{sleep}$ and $t_{awake}$ determine the rate (i.e. the Poisson distribution) that generates the events.
}
\label{fig:model_illustration}
\end{figure}

Fig.~\ref{fig:model_illustration} shows an illustration of the model idea.

We now propose four different models, which differ in the assumptions made on the relation of the rate and sleep/awake time parameters for different days.

\subsection*{Pooled-Pooled Model: Pooled Times and Rates}
The simplest model assumes that for a given user there is a single $\lambda_{awake}$; i.e. the user has very similar phone interaction patterns each day.
Also $t_{sleep}$ and $t_{awake}$ are each identical for all days, that is: the user goes to sleep, and wakes up, at the same times each day:

\begin{align*}
t_{sleep} &\sim \operatorname{DiscreteUniform}(0, n) \\
t_{wake} &\sim \operatorname{DiscreteUniform}(0, n) \\
\lambda_{awake} &\sim \operatorname{Gamma}(2.5, 1) \\
\end{align*}
Here $\operatorname{DiscreteUniform}(0, n)$ represents a uniform probability to choose a timebin between 0 and $n=96$.
No additional prior knowledge of $t_{sleep}$ and $t_{awake}$ is assumed; there is equal probability of any bin value. In other words, sleep and awake time are equally probable at any time of the day.
The prior for $\lambda_{awake}$ is chosen to represent our prior belief of a reasonable rate of events, specifically with both mean and variance = 2.5 (events/bin) and a longer tail than a normal distribution.

\subsection*{Independent-Pooled Model: Independent Times}
A somewhat more realistic model would assume that each day has independent $t_{sleep}$ and $t_{awake}$ times, while still sharing $\lambda_{awake}$ rates.
Therefore in this model there are $t_{sleep}^i$ and $t_{awake}^i$, with $i=1...m$, one for the each of the considered days:

\begin{align*}
t_{sleep}^i &\sim \operatorname{DiscreteUniform}(0, n) \text{ for } i=1...m \\
t_{wake}^i &\sim \operatorname{DiscreteUniform}(0, n) \text{ for } i=1...m \\
\lambda_{awake} &\sim \operatorname{Gamma}(2.5, 1) \\
\end{align*}
The rest of the model remains as above.

\subsection*{Independent-Independent Model: Independent Times and Rates}
It may further be assumed that each day could have its own specific activity rate.
We modeled this as separate $\lambda_{awake}^i$ for each of the $m$ days, in addition to $t_{sleep}$ and $t_{awake}$ for each of the $m$ days:

\begin{align*}
t_{sleep}^i &\sim \operatorname{DiscreteUniform}(0, n) \text{ for } i=1...m \\
t_{wake}^i &\sim \operatorname{DiscreteUniform}(0, n) \text{ for } i=1...m \\
\lambda_{awake}^i &\sim \operatorname{Gamma}(2.5, 1) \text{ for } i=1...m
\end{align*}

\subsection*{Independent-Hyper Model: Hierarchical Rates}
The assumption that each day's interaction rate is completely independent may not be correct.
It may not be unreasonable to imagine that the daily rate(s) arise from an underlying user-specific rate; i.e. the user may have certain habits that varies from day to day but share some similarities specific to that user.
This is modeled by adding $\alpha_{\lambda}$ and $\beta_{\lambda}$ hyperparameters to the Gamma priors for $\lambda_{awake}^i$:

\begin{align*}
t_{sleep}^i &\sim \operatorname{DiscreteUniform}(0, n) \text{ for } i=1...m \\
t_{wake}^i &\sim \operatorname{DiscreteUniform}(0, n) \text{ for } i=1...m \\
\alpha_{\lambda} &\sim \operatorname{Exponential}(1) \\
\beta_{\lambda} &\sim \operatorname{Exponential}(1) \\
\lambda_{awake}^i &\sim \operatorname{Gamma}(\alpha_{\lambda}, \beta_{\lambda}) \text{ for } i=1...m
\end{align*}
We do not have strong prior beliefs for $\alpha$ and $\beta$, so we set their prior distributions to generic exponential distribution with rate parameter = 1,  $\operatorname{Exponential}(1)$.

\subsection*{Hyper-Hyper Model: Hierarchical Times and Rates}
Finally we could assume that each day's sleep and awake times derive from an underlying circadian rhythm that is specific to the user, but still modulated by events that take place during the week.
This can be modeled by changing the $t_{sleep}^i$ and $t_{awake}^i$ priors to a normal distribution, with hyperparameters $\alpha_t$, $\beta_t$ and $\tau_t$ as follows:

\begin{align*}
\alpha_t &\sim \operatorname{Exponential}(1) \\
\beta_t &\sim \operatorname{Exponential}(1) \\
\tau_t &\sim \operatorname{Gamma}(\alpha_t, \beta_t) \\
t_{sleep}^i &\sim \operatorname{Normal}(8*(n/24), \tau_t) \text{ for } i=1...m \\
t_{wake}^i &\sim \operatorname{Normal}(15*(n/24), \tau_t) \text{ for } i=1...m \\
\alpha_{\lambda} &\sim \operatorname{Exponential}(1) \\
\beta_{\lambda} &\sim \operatorname{Exponential}(1) \\
\lambda_{awake}^i &\sim \operatorname{Gamma}(\alpha_{\lambda}, \beta_{\lambda}) \text{ for } i=1...m
\end{align*}
The $t_{sleep}^i$ are here chosen to be centered at the bin corresponding to 23:00, while the $t_{awake}^i$ are centered at the bin corresponding to 07:00.
Also in this case we have no strong prior knowledge of the $\tau_t$, $\alpha_t$ and $\beta_t$ parameters, so we set their prior distribution to a non-informative Exponential and Gamma respectively.

\subsection*{Model Fitting and Selection}
\label{section:fitting}
The models are fitted using Markov Chain Monte Carlo (MCMC) sampling~\cite{gelman2014bayesian}, where the parameter values are estimated by a random walk in the parameter space guided by the log likelihood.
We use the \emph{pymc3} python library \cite{patil2010pymc, pymc3gh} for running the sampling, but any MCMC framework could be used to implement our model.
The result of the Bayesian inference is a trace that captures the most probable values of the parameters, and also gives an indication of the uncertainty of the estimation.

It is important to note that the models are unsupervised, which means that they are fitted only to the number of events without having access to the ground truth of the actual sleep patterns.
This allows the model to be fit to other datasets where we do not have ground truth of sleep patterns, which is desirable if the sleep inference has to be deployed on a large scale.
For dataset A we verify the fit by comparing with the sleep patterns from sleep trackers, while for dataset B we evaluate the fit by inspecting the inferred sleep patterns.

In order to find the model that provides the best overall fit for the intended purpose without introducing too many degrees of freedom, we compare the log posterior from the traces of the models, $\operatorname{logp}$, and see how they converge.

\begin{figure}
\centering
\includegraphics[width=1\columnwidth]{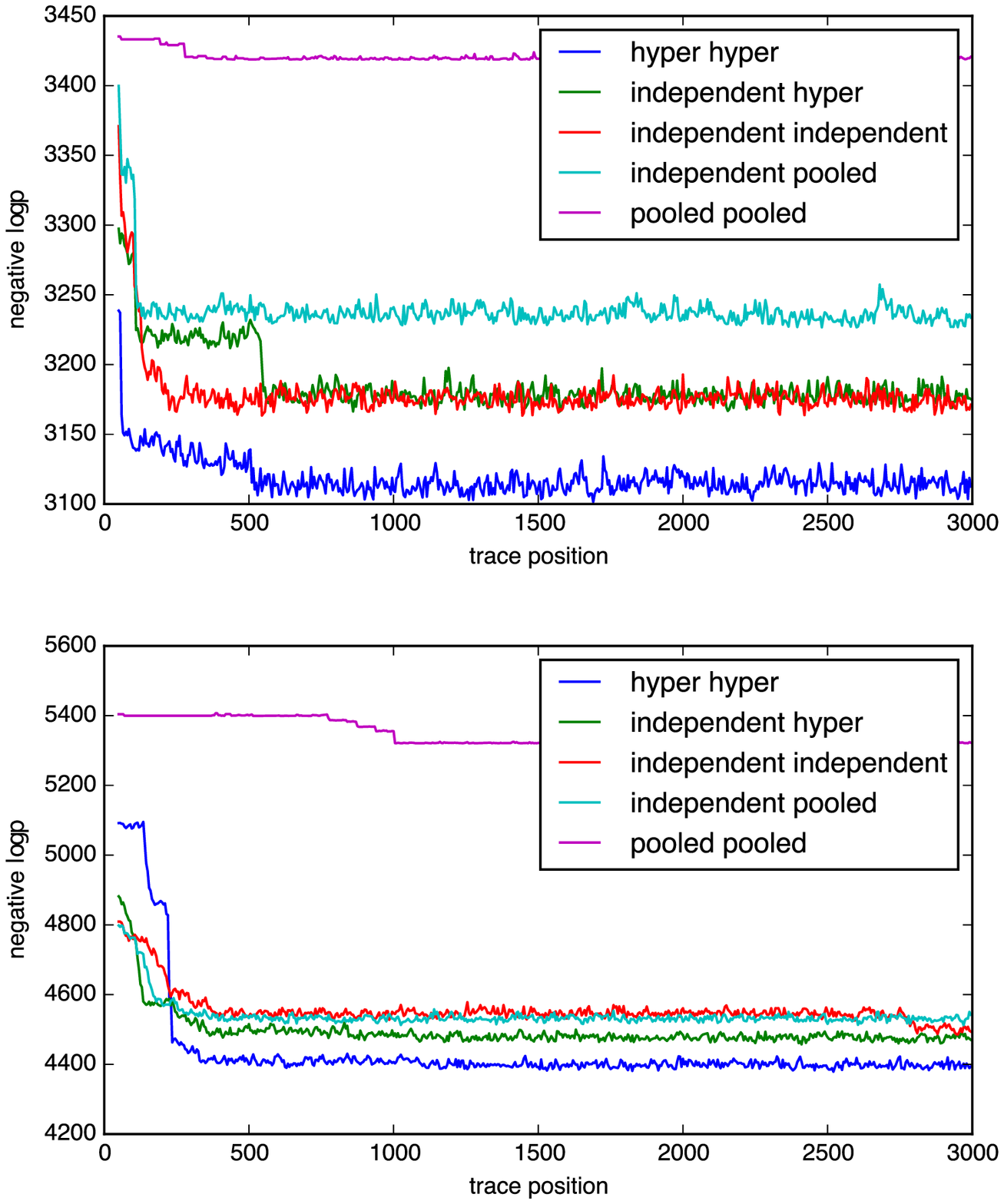}
\caption{Typical $\operatorname{logp}$ traces (A top, B bottom)}
\label{fig:logp}
\end{figure}

One example of a plot of $\operatorname{logp}$ traces for the five models is shown in Fig.~\ref{fig:logp}, which shows that the hyper-hyper model (blue) has the highest (least negative) $\operatorname{logp}$, followed by the independent-hyper model for dataset B. 
The three other models appear with lower $\operatorname{logp}$. 
In $76\%$ of the analyzed cases of dataset A ($84\%$ for dataset B), the hyper-hyper model has the highest $\operatorname{logp}$ score, followed by the independent-hyper model with the highest $\operatorname{logp}$ in $11\%$ ($13\%$) of the cases.

\begin{figure}[ht!]
\centering
\includegraphics[width=1\columnwidth]{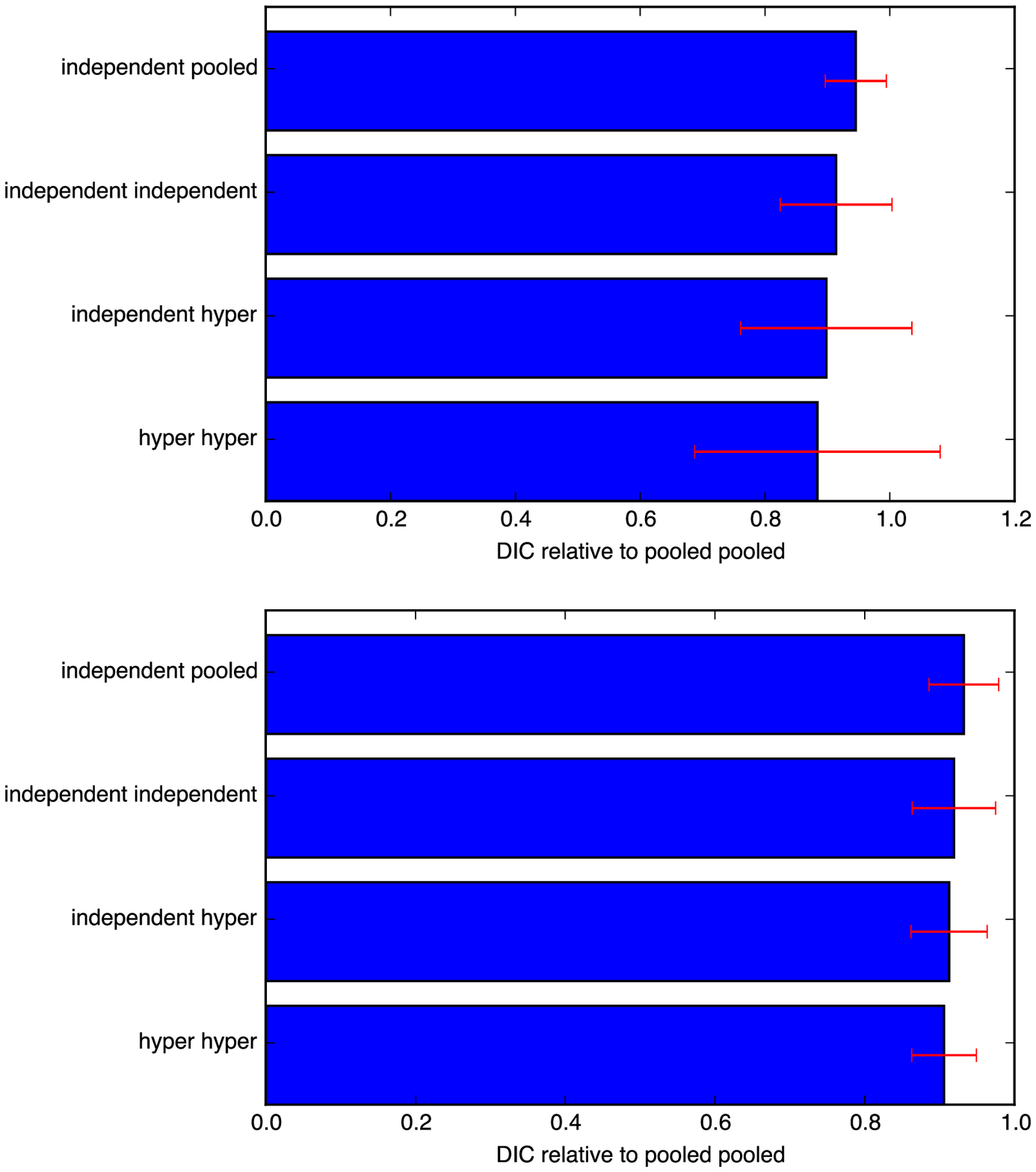}
\caption{Relative DIC scores (A top, B bottom), sorted by their mean value (error bars represent one standard deviation).
For both datasets the order is the same, with the hyper-hyper model having the lowest mean DIC.}
\label{fig:dicrel}
\end{figure}

The $\operatorname{logp}$ estimation does not, however, take into account the added complexity of the more advanced models. 
An attempt to do so is the Deviance Information Criterion ($\operatorname{DIC}$)~\cite{berg2004deviance}, which penalizes the increased degrees of freedom (more model parameters) that usually result in a model that is easier to fit to the data.
Fig.~\ref{fig:dicrel} shows the Relative DIC score (vs. the simplest model, pooled-pooled).
The order is identical for both datasets.

Further, Table~\ref{table:dic} compares the 5 models by ranking the calculated DIC for all $126$ and $324$ users. 
The \emph{median} rank shows that the hyper-hyper model is the ``best'' model; it has a probability of being the best ranked model (\emph{p(Best)}) in $62\%$ of the cases for dataset A ($69\%$ for dataset B).
The independent-hyper model follows as a somewhat distant 2nd best, ranking highest in $17\%$ ($19\%$) of the cases.

It should be noted that, in addition to their different abilities to reflect the underlying assumptions and provide varying levels of fit to the actual data, the models also differ in their runtime; the most complex model typically takes $15$ times longer to execute than the simplest. 
In particular, the hyper-hyper model on average had a runtime that is $60\%$ longer than the independent-hyper model, so there may be cases where the latter would be a better model to use despite the slightly worse DIC ranking.

\begin{table*}[]
\resizebox{1\textwidth}{!}{
\begin{tabular}{@{}|l|l|c|c|c|c|c|c|@{}} \hline
&Model Ranks    & Median & \multicolumn{2}{c|}{Mean} & p(Best) & \multicolumn{2}{c|}{Mean Relative DIC} \\ \hline
&                        &        & Value      & (StdDev)      &         & Value             & (StdDev)         \\ \hline
A&pooled-pooled           & 5      & 4.27       & (1.37)        & 0.10    & 0.96              & (0.16)           \\ \hline
 &independent-pooled      & 4      & 3.82       & (0.85)        & 0.03    & 0.95              & (0.05)           \\ \hline
 &independent-independent& 3      & 2.86       & (1.08)        & 0.08    & 0.91              & (0.09)           \\ \hline
 &independent-hyper       & 2      & 2.29       & (0.83)        & 0.17    & 0.90              & (0.14)           \\ \hline
 &hyper-hyper             & 1      & 1.76       & (1.11)        & 0.62    & 0.88              & (0.20)           \\ \hline
B&pooled-pooled           & 5      & 4.70       & (0.89)        & 0.02    & 0.99              & (0.01)           \\ \hline
 &independent-pooled      & 4      & 3.75       & (0.66)        & 0.02    & 0.93              & (0.05)           \\ \hline
 &independent-independent& 3      & 2.92       & (1.02)        & 0.09    & 0.92              & (0.06)           \\ \hline
 &independent-hyper       & 2      & 2.06       & (0.69)        & 0.19    & 0.91              & (0.05)           \\ \hline
 &hyper-hyper             & 1      & 1.56       & (0.94)        & 0.69    & 0.91              & (0.04)           \\ \hline
\end{tabular}
}
\caption{Model DIC comparisons}
\label{table:dic}
\end{table*}

\section*{Results}

All five models have been run on both datasets, producing an estimation of the times of sleep and wake up for each day, as well as estimates for the other hyperparameters, for each user.
Moreover, we calculated logp and DIC as discussed in the previous section.
We firstly verify the accuracy our method using the ground truth from the sleep trackers.
We then provide a qualitative analysis of some key examples of individual sleep patterns, and a description of the aggregated sleep patterns for both datasets.
For the remainder of the paper we restrict our analysis to the model with the best fit, the hyper-hyper model.

\subsection*{Comparison to Related Work and to Ground Truth}

To assess the results, we compare the sleep periods inferred by our model and those inferred by a previously suggested rule-based method to the ground truth collected by the Sony sleep trackers.

For each day we calculate the time of sleep and time of awake inferred by our model as the mean of the $t^i_{sleep}$ and $t^i_{wake}$ respectively, and we consider the user asleep (Z~=~1) for all time bins between $t^i_{sleep}$ and $t^i_{wake}$, and awake (Z~=~0) for the remaining bins.

For a representative and comparable method, we chose to implement a rule-based algorithm similar to what is proposed by Abdullah et. al.~\cite{abdullah2014towards} to derive sleep data for dataset A. 
This rule-based method essentially works by finding the longest contiguous sleep period, with a prior assumption that sleep must start after 10 PM and before 7 AM next morning.
Note that the original algorithm is based on screen on-off events and furthermore discards events of short duration during the night; in our case we use app launches with no available duration, and thus cannot discard events of short duration.

For the sleep trackers we can directly mark each time bin as sleep (Z~=~1) if the trackers have detected at least one sleep status in that bin, and awake (Z~=~0) otherwise.

We again consider one user at a time.
For each user we now have three binary matrices: two inferred sleep status values per time bin from either model, and one measured sleep status value per time bin (ground truth) .
We evaluate this as two binary classification problems, and calculate accuracy, precision, recall and F1 for each model and for each user according to the definitions:

\[
 \operatorname{accuracy} = \frac{\operatorname{correct~predictions}}{\operatorname{predictions}}
\]

\[
 \operatorname{precision} = \frac{\operatorname{true~positives}}{\operatorname{predicted~positives}}
\]

\[
 \operatorname{recall} = \frac{\operatorname{true~positives}}{\operatorname{all~positives}}
\]

\[
 \operatorname{F1} = 2 \cdot \frac{\operatorname{precision}\cdot\operatorname{recall}}{\operatorname{precision}+\operatorname{recall}}
\]

\begin{figure*}[ht!]
\centering
\includegraphics[width=1\columnwidth]{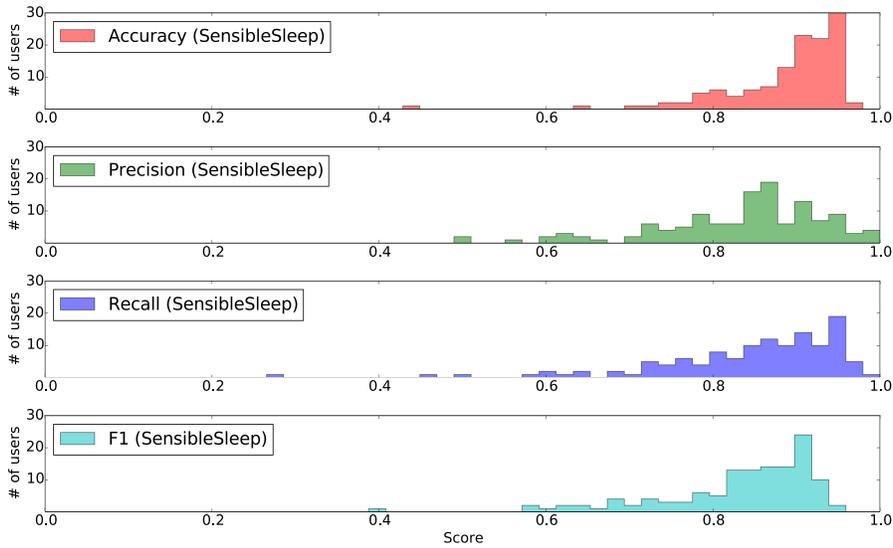}
\caption{Histogram of the calculated accuracy, precision, recall and F1 score for users in dataset A, comparing the proposed method to the sleep tracker ground truth.}
\label{fig:mcmc_acc}
\end{figure*}

\noindent Fig.~\ref{fig:mcmc_acc} shows the resulting distribution of accuracy, precision, recall and F1 scores for the proposed method.
The SensibleSleep method achieves a mean accuracy of 0.89, and a mean F1 score of 0.83.
The below-average scores for some users are expected, since it is likely that among the large population under study there will be people having irregular sleep schedule or noisy sleep ground truth.

\begin{figure*}[ht!]
\centering
\includegraphics[width=1\columnwidth]{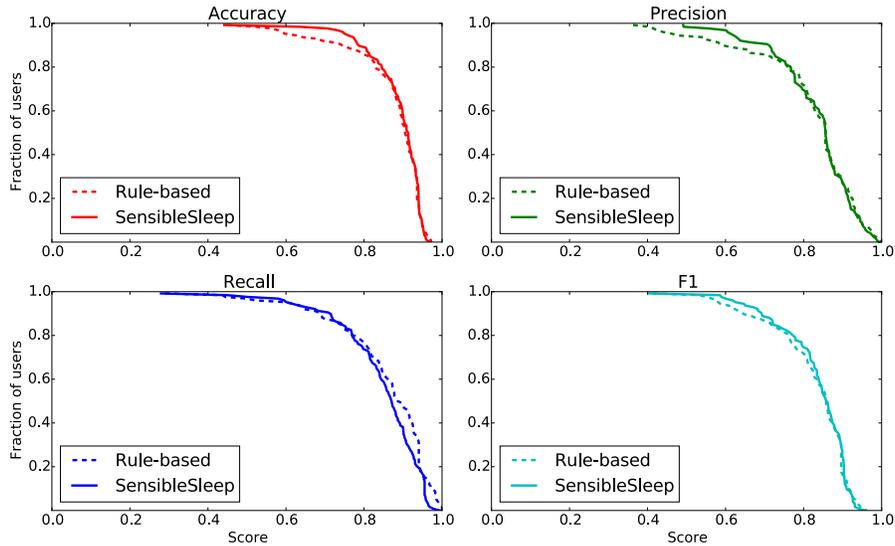}
\caption{Complementary cumulative distribution of accuracy, precision, recall and F1 scores for users in dataset A, comparing the proposed model (solid line) to the rule-based model (dashed line), showing the proportion of users (y-axis) having a score less than or equal to a specific value (x-axis).}
\label{fig:model_compare_acc}
\end{figure*}

Fig.~\ref{fig:model_compare_acc} shows the corresponding \textit{complementary cumulative distributions} of the accuracy, precision, recall and F1 scores of the proposed SensibleSleep model vs that of the rule-based model~\cite{abdullah2014towards}.
The results are generally comparable between the two models, on this particular dataset.
Our model has slightly better accuracy and precision whereas the previously suggested rule-based model has a slightly better recall. 
The F1 scores, which weights precision and recall equally, are comparable.
This particular dataset has only very limited sleep interruptions during the night.
For populations with more interrupted sleep, we expect our model to maintain a high score.

\subsection*{Individual Sleep Patterns}
We now analyze individual sleep patterns to show the results of the model in details.
For each user we create a visualization of sleep schedules. We call this the \emph{sleep matrix}.
Each row represents one day, and each column represents one time bin.
The blue color shows the probability that sleep takes place within the interval; the darker the color the higher the probability.
The red dots show activity count per bin; the larger the radius the more events are registered within that particular bin.
This compact representation is able to capture at a glance the sleep patterns of individuals over time.
We have created one such sleep matrix for each of the users, which allows us to inspect hundreds of sleep patterns quickly.
Large individual variability both in sleep schedules (regular, irregular) and in phone activity (low, high, during day or night) are noticeable.
Still, in most cases it is evident that the model is able to capture a reasonable sleep period, even if it may have been somewhat interrupted.

\begin{figure*}[ht!]
\centering
\includegraphics[width=1\columnwidth]{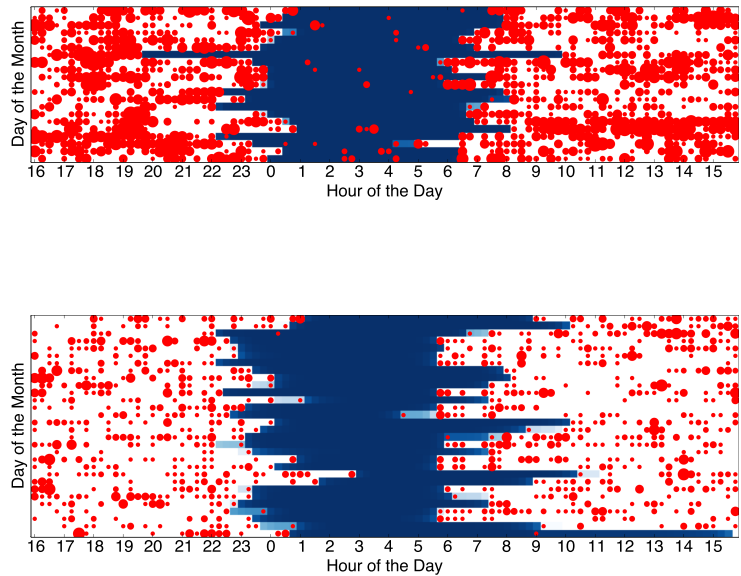}
\caption{Sleep matrix of two sample users (21 days from dataset A top, 30 days from dataset B bottom)}
\label{fig:grid}
\end{figure*}

Let us consider the inferred sleep patterns for two example users in Fig.~\ref{fig:grid}. 
The top user has a pretty regular schedule, waking up around 5:30 except every few days, when he/she wakes up later -- presumably due to vacation or weekends.
Notice the light blue sections that indicate how the model is less confident about the probability of sleep due to events that do not follow the usual patterns.
The bottom user instead has a much more unstable app usage, therefore the model infers a correspondingly more unstable sleep schedule.
The bottom user has also some events in the middle of the night throughout many days (which is presumably checking the phone at night) yet the model is still able to correctly infer this being a sleep phase.
Finally notice how the two users have significantly different intensity of app usage (the bottom one uses the phone much more than the top one), yet this is not a problem since the model learns individual activity rates.

\newpage

\subsection*{Aggregated Sleep Schedules}
In this section we also quantify the aggregated sleep patterns.
From the posterior probability distribution functions ($\operatorname{PDF}$s), $P_{t_{sleep}}(t)$ and $P_{t_{awake}}(t)$, the probability that the user is sleeping can be estimated as follows:

\begin{align*}
P_{sleep}(t) = P_{t_{sleep}}(t) - P_{t_{awake}}(t)
\end{align*}

This is equivalent to stating that a user is currently sleeping if he has passed the time of falling asleep but has not yet passed the time waking up.

\begin{figure}
\centering
\includegraphics[width=1\columnwidth]{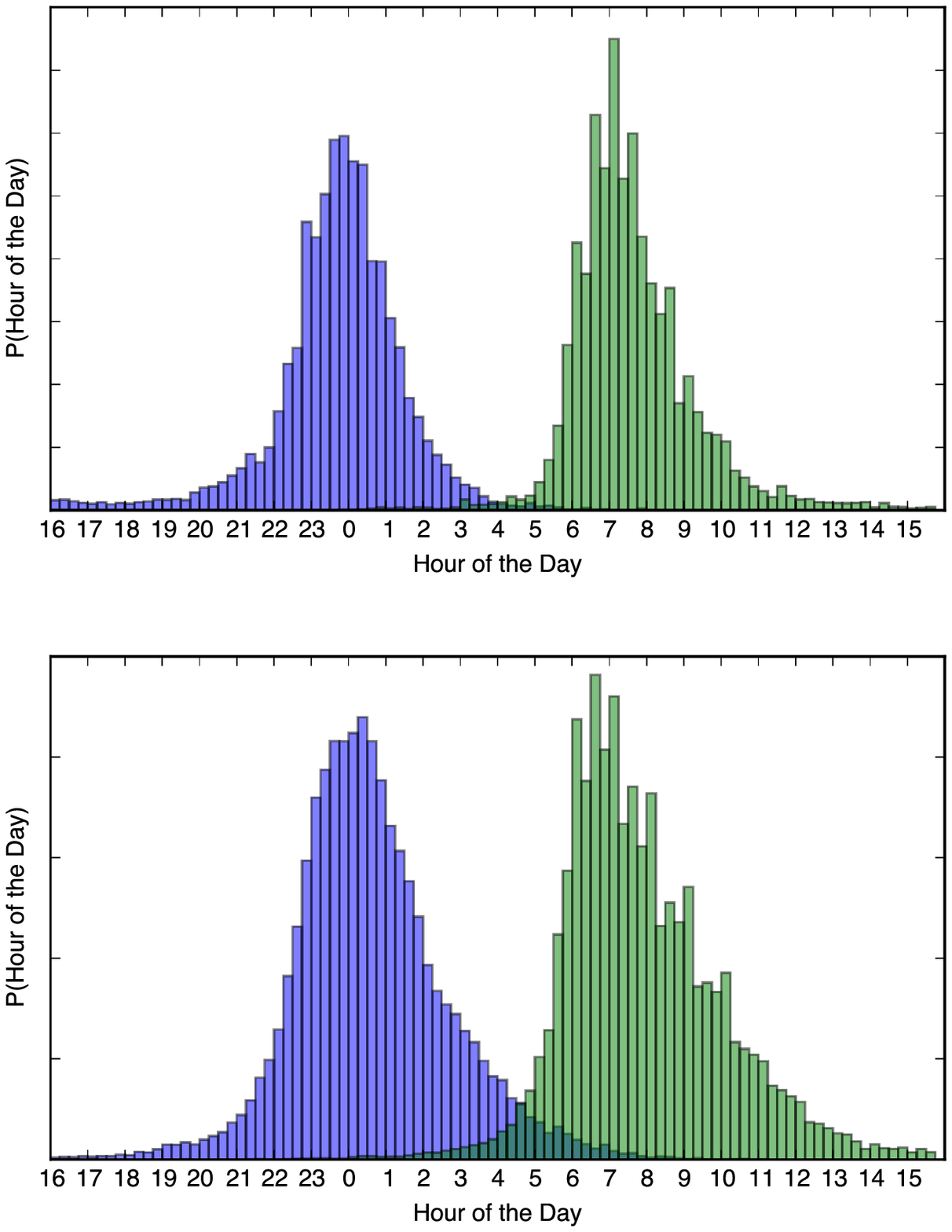}
\caption{Aggregate Posterior Probability Distributions of $t_{sleep}$ (blue) and $t_{awake}$ (green) (A top, B bottom), showing what the probability is for the specific population to go to sleep or wake up at the specified time.}
\label{fig:tpdf}
\end{figure}

The derived values of sleep-length $t_{sleeplength}$ and mid-sleep time $t_{midsleep}$ can be calculated directly from the values of $t_{sleep}$ and $t_{awake}$ for each sample of the trace, and the posterior density can be estimated for these derived values in a similar way as for the model parameters.
Fig.~\ref{fig:tpdf} shows the aggregate posterior probability density functions for $t_{sleep}$ and $t_{awake}$ for the $126$ users of dataset A over $15-30$ days, and for the $324$ users of dataset B over a selected period of $30$ days (just after semester start).
It may not be entirely meaningful to average the sleep patterns from all users, but it serves to illustrate the distribution of $t_{sleep}$ and $t_{awake}$ for a larger population.
Table~\ref{table:sleeptimes} summarizes the sleep and wake times.

\begin{table}[]
\centering
\begin{tabular}{@{}|l|c|c|c|c|@{}}
\hline
  & \multicolumn{2}{c|}{Sleep Time} & \multicolumn{2}{c|}{Wake Time} \\ \hline
  & Mean          & (Std)          & Mean         & (Std)          \\ \hline
A & 23:38         & (2h 16m)         & 7:40         & (2h 2m)         \\ \hline
B & 0:35          & (2h 6m)         & 7:55         & (2h 15m)         \\ \hline
\end{tabular}
\caption{Aggregated sleep and wake times}
\label{table:sleeptimes}
\end{table}

\begin{figure}
\centering
\includegraphics[width=1\columnwidth]{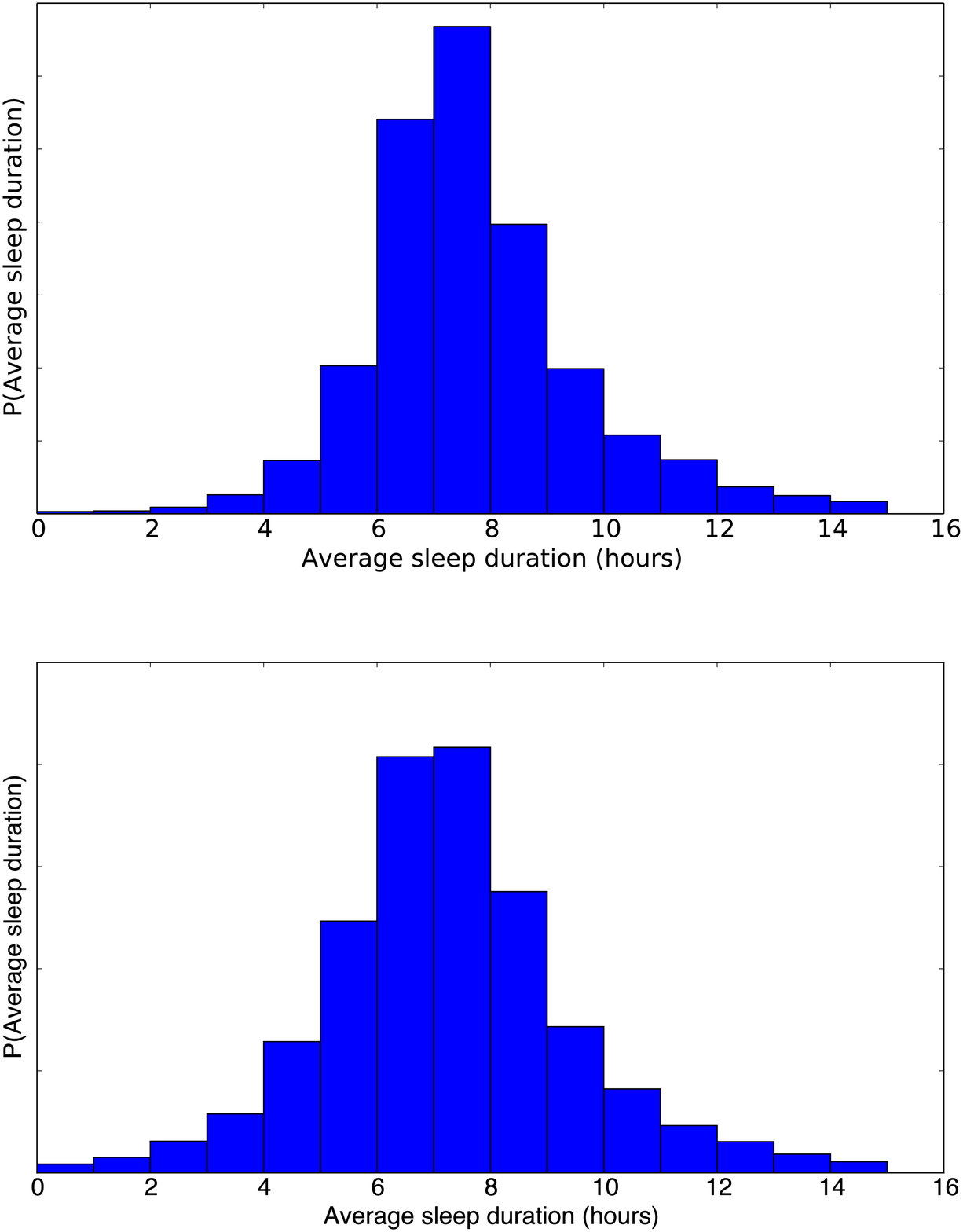}
\caption{Aggregated Sleep Durations (A top, B bottom), based on the Posterior Probability Functions. This illustrates the probability of the length of a nights sleep within the population within the datasets.}
\label{fig:sleepdur}
\end{figure}

Across the 30 (14-28) analyzed days for the $324$ ($126$) users of the study, the distribution of sleep durations are as shown in Fig~\ref{fig:sleepdur}. %
The model allows us to easily compute such metrics. 
The mean value is around 8:02 ($\pm$2h 36m) for dataset A and 7:20 ($\pm$2h 28m) for dataset B. 
Notice how the distributions are not completely similar; this is likely due to the fact that the larger dataset B captures the sleeping behavior of students as opposed to dataset A that may have a more diverse demographic distribution.

\begin{figure}
\centering
\includegraphics[width=1\columnwidth]{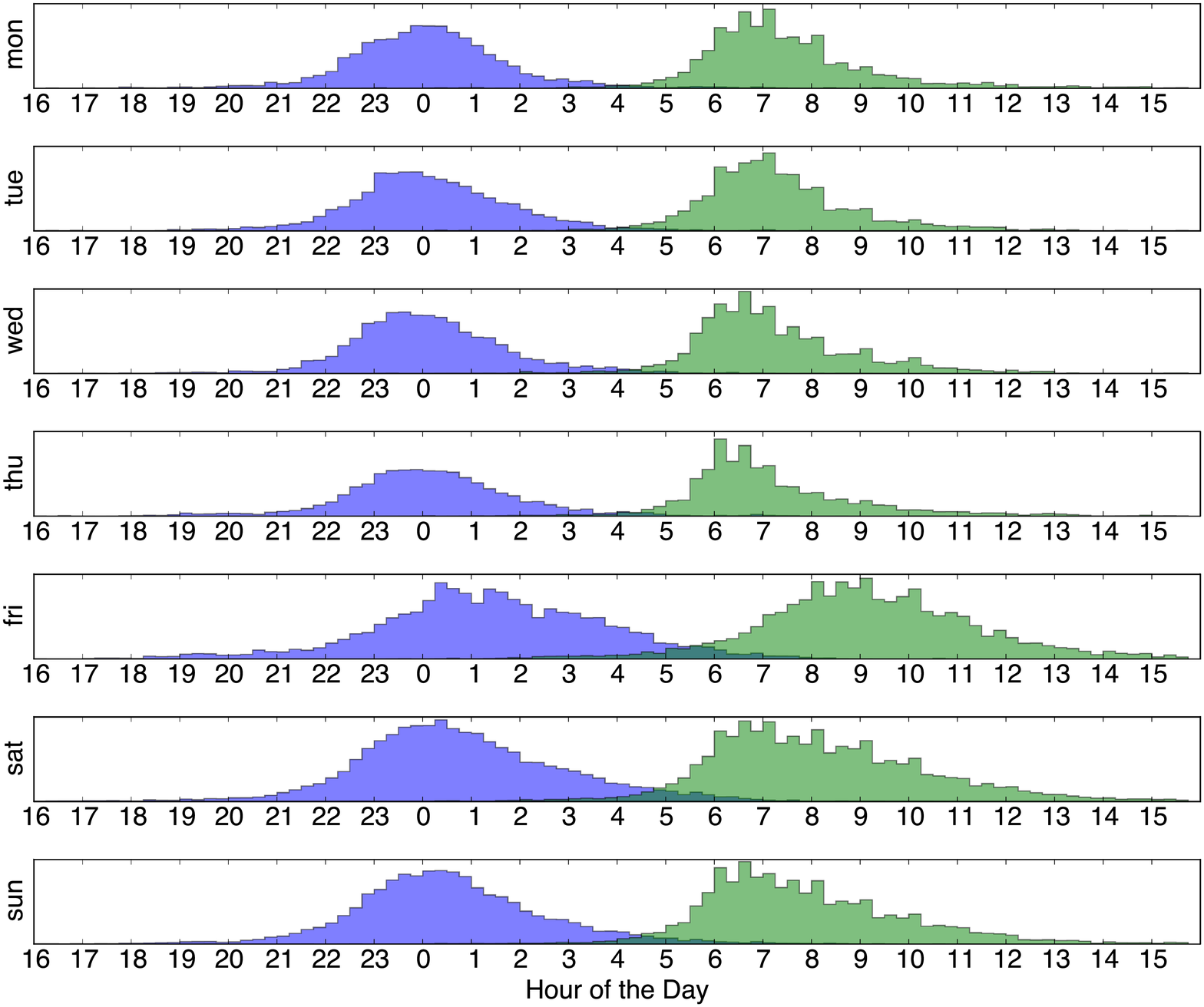}
\caption{$t_{sleep}$ (blue) and $t_{awake}$ (green) over weekdays for dataset B}
\label{fig:weekdays}
\end{figure}

Fig.~\ref{fig:weekdays} shows the probability density functions for the $t_{sleep}$ and $t_{awake}$ times for all users of dataset B, grouped according to weekday.
Mondays to Thursdays appear quite similar, but Friday shows a much wider distribution; users typically go to bed much later on Friday and sleep in on Saturday. The distributions start to narrow down Saturday and Sunday but are more ``week-like'' only from Tuesday morning again.

\newpage

\section*{Discussion}
The main contribution of this work is to show how simple counts of smartphone interactions can be used to infer sleep patterns with reasonably high accuracy.
We have demonstrated how the seemingly weak signal of screen events carry significant information of the user status.
Our method has several advantages:
\begin{itemize}
\item The method requires only a smartphone and can therefore be deployed without the need for special equipment or methods, such as fitness or sleep tracking bands, or sleep diaries.
\item The data collection is completely automated, as no action is required from the user in setting up the tracking or remembering to log his/her activity. 
\item Since the model requires only screen interactions, it is absolutely non-intrusive and privacy-preserving. 
Although in this work we stored the data on a central server for analysis purposes, the data could remain on the phones and the sleep analysis could in principle be run directly on the phones as well.
\item Compared to accelerometer or microphone-based methods, using only screen events is much more battery-efficient.
\end{itemize}

Although solutions using screen events have been proposed before~\cite{abdullah2014towards, Jayarajah:2015:CCY:2800835.2804393}, our model provides a number of key improvements:
\begin{itemize}
\item It is more robust to noise such as screen events generated by checking the phone at night.
\item Using a Bayesian formulation allows us to provide confidence intervals for the sleep and awake times, instead of point estimates only.
\item It does not depend on ad-hoc rules, but it is based on a well-defined statistical formulation.
\item It is fitted and verified on a much larger userbase of over 400 users, and a longer time duration (between 2 and 4 weeks).
\end{itemize}

Demonstrating the feasibility of inferring reasonable sleep patterns from simple event counts opens the way for new exciting research directions.
In particular we believe that similar methods can be applied to large datasets of user activity.
For example on social network (such as Twitter, Facebook, Meetup, Gowalla) users leave a trace of their activity in the form of messages, posts, likes, etc.
Another great example is Call Detail Records, the logging information kept by telecom providers about user calls and SMS.
These events could be treated again as a proxy for sleep and wake cycles.

The main drawback of the proposed method is that it requires that users periodically interact with their phones during their wake time.
In line with other recent polls (see for example\cite{tecmark, gallup, bank-of-america}), we show that in most cases this does happen, as the population of users analyzed here tend to check their phone from the early morning to the late night when awake.
Different populations, however, such as elderly people less accustomed to smartphone usage, may not show similar usage patterns.
There is therefore a need for additional work in order to understand how increased sparsity would affect sleep pattern reconstruction.

\section*{Conclusions}
We have presented a Bayesian model to infer sleep patterns from smartphone interactions, which we have applied to two datasets of more than 400 users in total.
We have compared the model output with ground truth from sleep trackers, and we have shown how the model is able to recover the sleep state with a mean accuracy of 0.89 and a mean F1 score of 0.83.
Furthermore, we have shown how the model is capable of producing very reasonable individual and aggregated sleep patterns.
Our method represents a cost-effective, non-intrusive and automatic alternative for inferring sleep patterns, and can pave the way for large-scale studies of sleep rhythms.

\section*{Acknowledgments}

AC is funded in part by the High Resolution Networks project (The Villum Foundation), as well as Social Fabric (University of Copenhagen).
PB is supported in part by the Innovation Fund Denmark through the project Eye Tracking for Mobile Devices.
VS and HJ received funding from Sony Mobile. 
The funders had no role in study design, data collection and analysis, decision to publish, or preparation of the manuscript.

\end{document}